\documentclass[12pt,preprint]{aastex}

\raggedright

\usepackage{natbib}
\shorttitle{Contact Binary Variables as X-ray Sources}
\shortauthors{Chen et al.}

\begin{document}
   
\title {Contact Binary Variables as X-ray Sources }

\author{W. P. Chen, Kaushar Sanchawala}
 \affil{Graduate Institute of Astronomy, National Central University, 
        Chung-Li, Taiwan}
 \email{wchen@astro.ncu.edu.tw}

\and

\author{M. C. Chiou}
 \affil{Department of Physics, National Central University, Chung-Li, Taiwan}

\begin{abstract}
We present cross-identification of archived X-ray point sources with 
W~UMa variable stars found in the All-Sky Automated Survey (ASAS). 
In a surveyed sky area of 300 square degrees of ASAS, 36 W~UMa stars 
have been found associated with X-ray emission.  We compute the distances 
of these W~UMa systems and hence their X-ray luminosities.  
Our data support the "supersaturation" phenomenon seen in these fast rotators, 
namely that the faster a W~UMa star rotates, the weaker its 
X-ray luminosity.     
\end{abstract}


%%%%%%%%%%%%%%%%%%%%%%%%%%%%%%%%%%%%%%%%%%%%%%%%%%%%%  
\section{W Ursae Majoris Stars in the ASAS Database}
%%%%%%%%%%%%%%%%%%%%%%%%%%%%%%%%%%%%%%%%%%%%%%%%%%%%%

W~Ursae Majoris (W~UMa) variables, also called EW stars, are contact 
eclipsing binaries of main-sequence component stars with periods 
$\sim 0.2$--1.4~d.  The component stars may have different masses, but  
fill their Roche lobes to share a common envelope.  Apparently because of an 
efficient convective mechanism to redistribute energy, the details of 
which are still unknown, both component stars have very similar brightness 
and chromospheric activity \citep{vil87}.  The light curves of W~UMa 
systems are therefore characterized by two nearly equal minima 
with virtually no plateau.  

W~UMa stars are known X-ray emitters.  \cite{ste01} examined a sample 
of some 100 such systems and found about half of them to be X-ray sources.  
The detailed X-ray emission mechanism in these contact systems remains elusive 
but is thought to be related to stellar dynamo magnetic activity, arising 
from a shared convective envelope plus synchronous, fast rotation \citep{gon04}.  
The X-ray emission of W~UMa stars is known to vary, especially during flares 
\citep{mcg96}.  Interestingly, monitoring of EUV variation of the 44i Bootis W~UMa system
by \citet{bri98} shows a period twice as long as that derived from optical light
curves.  This suggests that the high-energy photons come mainly from one of the
binary components, rather than from both.   An expanded sample of W~UMa systems 
with X-ray emission would be an important step to shed light 
on their X-ray nature, for example on possible correlation with binary 
evolution \citep{cru84} and with stellar rotation.  In this short communication 
we present our analysis of the X-ray emission of 34 W~UMa stars.   

Our sample of W~UMa stars was taken from the All-Sky Automated 
Survey (ASAS) database.  The ASAS project\footnote{
http://archive.princeton.edu/asas/}, sited at Las Campanas 
Observatory, started with a prototype ASAS$-$1 and ASAS$-$2 instruments 
with a $ 768\times 512$ Kodak CCD and 135/f1.8 telephoto lens to monitor 
stars brighter than I$\sim$~14 magnitudes \citep{poj97, poj00}.
From 1997 to 2000, more than 140,000 stars were observed in selected 
fields covering $\sim 300$ square degrees for nearly 50 
million photometric measurements.  This resulted in discovery of 
more than 3500 variable stars by ASAS$-$2, among which 380 are periodic 
variables.  The ASAS$-$3 system has been in operation since August 
2000, and up to 2002 had discovered over 1000 eclipsing binaries, almost 
1000 periodic pulsating variables, and over 1000 irregular stars 
among the 1,300,000 stars in the RA=0h-6h quarter of the 
southern sky \citep{poj02}.  The ASAS$-$3 database now includes about  
4000 entries in the list of contact eclipsing binaries (called EC in the 
ASAS classification).  Light curves for variables identified by 
ASAS$-$2 are available online without classification of variable types.  
For these, we inspected 380 periodic variables and identified 36 W~UMa 
candidates on the basis of their light curves.  The ASAS$-$3 database already 
categorizes variable stars and we took all entries in the EC class as of 2002  
as our W~UMa sample.   

We then searched for X-ray counterparts for our sample W~UMa stars in the  
$ROSAT$ All-Sky Survey database.  One should note that the light curves of 
W~UMa stars resemble those of RRc variables, a subclass of RR\,Lyr stars
\citep{hof85}.  RRc stars do have a different period range\footnote{Being 
binary, a W~UMa star would have two minima within a true period, hence with  
a light curve resembling that of a RRc star of half the peroid.}  and 
are not known to emit X rays.  In the course of our study, as byproduct 
we have identified 
2 Cepheids (ASAS\,052020$-$6902.4, ASAS\,234131$+$0126.4), 
4 $\beta$\,Lyrae variables (ASAS\,124435$-$6331.7, 
ASAS\,180253$-$2409.6, ASAS\,104006$-$5155.0, ASAS\,050527$-$6743.2), 
and 2 Algol-type variables (ASAS\,205603$-$1710.9, ASAS\,144245$-$0039.9)    
as possible X-ray sources.  For ASAS stars, in addition to visual inspection 
of their light curves, we also analyzed the power spectra by Fourier Transform 
to get check the periodicity, as a caution of using any database with pipeline 
analysis and classification.  As an illustration, an initial inspection 
of ASAS\,015647$-$0021.2 suggest 
a possible identification with a W~UMa or an RRc, with a period of  
$\sim 0.3511$~d.  The power spectrum shows the maximum signal has 
a period at 0.5427~d instead.  Because this variable star has an X-ray 
counterpart, we conclude it is a W~UMa star.  

At the end a total of 36 W~UMa stars, 7 in ASAS$-$2  
and 29 in ASAS$-$3, have been found with possible X-ray counterparts.  
In \S 2 we present the derivation of the 
X-ray luminosities of these W~UMa stars.  In \S we discuss how their X-ray 
luminosities correlate with rotational periods.      

%%%%%%%%%%%%%%%%%%%%%%%%%%%%%%%%%%%%%%%%%%%%%%%%%%%%%%%
\section{Distance and X-ray Luminosity Determinations}
%%%%%%%%%%%%%%%%%%%%%%%%%%%%%%%%%%%%%%%%%%%%%%%%%%%%%%%

We adopted the calibration scheme by \citet{ruc97b} to compute the absolute 
magnitude of each W~UMa star and hence its distance; namely,   

  $$M_{\rm V} = -4.42 \, \log P + 3.08 (B-V)_{0} + 0.10, $$

where the coefficients are the median values in the analysis of \citet{ruc97b}.  
Recently \citet{ruc04} refined this period-luminosity-color relation 
% as $$ M_{\rm V} = -4.44\, \log P + 3.02 (B-V)_{0} + 0.12, $$
with minor modifications of the coefficients.  The revised relation does not 
affect our results.   

% Two of the W~UMa stars with X-ray counterparts have periods or colors
% outside the ranges of the calibration, so were excluded from the 
% final list to compute the I-band absolute magnitudes and distances.

The $ROSAT$ All-Sky Survey (RASS) was carried out during 1990--1991 as a part of the 
$ROSAT$ mnission \citep{tru91}.  The survey was conducted in the 
soft X-ray (0.1--2.4~keV)
and in the extreme ultraviolet (0.025--0.2~keV) bands \citep{wel90}.  We searched 
the RASS database for each W~UMa stars in the ASAS catalog to find possible X-ray 
counterparts within a positional coincidence radius of 30\arcsec.  In almost every  
case, with only two exceptions,  
the cross-identification is relatively straightforward because either the 
nominal positions of the X-ray and optical sources match well, or no alternative 
star seems evident near the X-ray position. 

Our derivation of the X-ray luminosity followed that of \citet{ste01}.  
For each X-ray source, the source counts in the hard ($H$; 0.5--2.0~keV) and soft
($S$; 0.1--0.4~keV) bands are retrieved from the database and the hardness ratio
$HR=(H-S)/(H+S)$ was computed.  The X-ray luminosity for each source then was 
estimated by multiplying its $ROSAT$ X-ray count by an energy conversion factor, 
$ECF = (5.3 \, HR + 8.7) 10^{-12}$~erg~cm$^{-2}$~cts$^{-1}$ \citep{hun96}.

Our X-ray W~UMa stars are listed in Table~\ref{tab:1}.  The first columns give 
the running numbers.  The next two columns give the coordinates for each star, 
followed by its period (taken
from ASAS database), and the observed $(B-V)$ color.  The V magnitude was taken 
from the ASAS database and the B magnitude was from SIMBAD, which for our sample, 
was found typically to be within 0.02~mag of the USNO B magnitude.  Columns 6 
and 7 list respectively the derived distance of the W~UMa star and its X-ray luminosity.
The last columns give the known star identification, taken from SIMBAD, with known 
W~UMa stars marked with "WU".  
DY~Cet is known to be a W~UMa star from {\it HIPPARCOS} light curve \citep{sel04}.
The two sources with the counterparts outside of 
the 30\arcsec search radius are each 
labeled with an asterisk.  
% Of all the entries in Table~\ref{tab:1} 
% only star No.~1 has been known previously as an X-ray W~UMa star.  
Three of the brightest stars in our sample have parallax measurements by {\it HIPPARCOS}, with 
which the inferred distances can be compared.  The results are shown in Table~\ref{tab:2}. 

%Note that
%the two "outliers", with periods longer than 0.6~d, are the hottest
%stars in our sample.

% $$ M_{\rm V} = -4.44\, \log P + 3.02 (B-V)_{0} + 0.12, $$

% DY Cet was found to be a W~UMa star from Hipparcos light curve
%  (Selam, 2004, AA, 416, 1097)
%  P=0.4408
%  simbad B=10.4, V=9.8  B-V=0.6,  Mv=3.5, d=182 pc
%  our data              B-V=0.93, Mv=4.51, d=114 pc
%  USNO NOMAD B=9.987, V=9.595, R=9.340, B-V=0.39 --> Mv=2.88, d=220 pc

% YY Eri given by Rucinski 1997b, P=0.3215  B-V=0.66 Mv=4.44 +/- 0.14
%  simbad B=9.05, V=8.41  B-V=0.64
%  Rucinski relation --> Mv=4.3,  d=66 pc
%  Our data B-V=0.68 --> Mv=4.36, d=64 pc

% HN Eri
% P=0.6236
% Simbad B=8.91, V=8.68  B-V=0.23 --> Mv=1.72, d=247 pc
% ours                   B-V=0.29 --> Mv=1.91, d=226 pc

\begin{deluxetable}{rrrrrrrl}
\tablewidth{0pt}
\renewcommand{\arraystretch}{.6}
%\tabletypesize\footnotesize
\tablenum{1}
\pagestyle{empty}
\tablecaption{Parameters of X-ray W~UMa Stars}
\tablehead{
 \colhead{No.} & \colhead{RA} & \colhead{DEC} & \colhead{P} & \colhead{(B-V)}
                & \colhead{Dist.} & \colhead{L$_{\rm X}$} & \colhead{Remarks}\\
 \colhead{}    & \colhead{h~~m~~s} &\colhead{ \arcdeg~~ \arcmin~~ \arcsec} 
                                            & \colhead{(d)} & \colhead{(mag)} 
                & \colhead{ (pc)} & \colhead{erg\,s$^{-1}$} & \colhead{}
          } 
\startdata
 1 & 00:14:47 & -39:14:36 & 0.3644 &  0.63 & 389 & 5.49 (30) & UY Scl \\
 2 & 00:17:21 & -71:55:00 & 0.5948 &  0.38 & 377 & 2.26 (31) & AQ Tuc \\ 
 3 & 00:23:28 & -20:41:48 & 0.4147 &  0.60 & 185 & 2.45 (30) & HD\,1922 \\
 4 & 00:42:40 & -29:56:42 & 0.3017 &  0.77 & 192 & 2.40 (30) & -\\
 5 & 01:37:11 & -34:59:18 & 0.4643 &  0.68 & 316 & 4.75 (30) & BV\,917 \\
 6 & 01:48:54 & -20:53:36 & 0.3169 &  0.80 & 143 & 1.48 (30) & TW\,Cet \\
 7 & 01:56:47 & -00:21.2  & 0.3514 &  0.97 & 384 & 3.96 (30) & \\ % 
 8 & 02:01:19 & -37:04:54 & 0.3081 &  1.06 &  77 & 7.86 (29) & XZ\,For \\
 9 & 02:29:02 & -30:25:48 & 0.3865 &  0.46 & 227 & 1.62 (30) & HD\,15517 \\
10 & 02:38:33 & -14:17:54 & 0.4408 &  0.93 & 107 & 7.29 (29) & DY\,Cet; WU \\
11 & 03:27:36 & -72:50:54 & 0.3099 &  0.76 & 157 & 1.76 (30) & CPD-73\,219 \\
12 & 03:37:02 & -41:31:42 & 0.2923 &  0.81 &  90 & 1.17 (30) & FX\,Eri; WU \\
13 & 03:44:44 & -61:05:48 & 0.2848 &  0.49 & 129 & 2.12 (30) & HD\,23816 \\
14 & 03:48:09 & -58:39:48 & 0.4292 &  0.75 & 209 & 2.27 (30) & - \\
15 & 03:52:00 & -21:55:48 & 0.3352 &  0.52 & 249 & 2.25 (30) & BD-22\,694 \\  
16 & 03:58:51 & -51:10:36 & 0.3106 &  0.40 & 154 & 2.57 (30) & CCDM\,J03589-5111AB \\ 
17 & 04:10:37 & -38:55:42 & 0.4269 &  0.45 & 319 & 3.79 (30) & CD-39\,1360 \\ 
18 & 04:12:09 & -10:28:12 & 0.3215 &  0.68 &  67 & 1.45 (30) & YY\,Eri; WU \\
19 & 04:21:03 & -26:29:30 & 0.3959 &  0.36 & 259 & 1.57 (30) & CD-26\,1640 \\
20 & 04:25:59 & -21:29:00 & 0.3319 &  0.69 & 199 & 1.40 (30) & AN 626.1935 \\
21 & 04:33:25 & -23:56:18 & 0.6236 &  0.29 & 228 & 8.26 (29) & HN\,Eri \\
22 & 05:05:37 & -57:55:36 & 0.5578 &  0.92 & 339 & 1.88 (31) & [FS2003] 0241 \\ 
23 & 05:06:17 & -20:07:48 & 0.4486 &  0.57 & 172 & 6.45 (29) & BV\,996 \\
24 & 05:11:14 & -08:33:24 & 0.4234 &  0.64 & 140 & 6.68 (29) & ER\,Ori \\
25 & 05:18:33 & -68:13:32 & 0.2854 & 1.044 & 187 & 1.80 (30) & RW\,Dor; WU \\ % ASAS2
26 & 05:22:14 & -71:56:18 & 0.7766 &  0.33 & 687 & 1.07 (30) & XY\,Men \\
27 & 05:24:52 & -28:09:12 & 0.2758 &  0.64 & 171 & 1.14 (30) & CD-28\,2151 \\
28 & 05:28:32 & -68:36.2  & 0.4474 & 0.789 & 285 & 5.16 (29)  & HD\,269602 \\ %ASAS2
29 & 05:39:59 & -68:28:41 & 0.3622 & 0.746 & 296 & 5.18 (29) & HD\,269960; WU \\ %ASAS2
30 & 05:55:01 & -72:41:36 & 0.3438 &  0.77 & 152 & 1.02 (30) & BV\,435 \\
31 & $^{*}$11:47:37 & -63:10.5 & 0.3392 & 1.131 & 257 & 2.24 (30) & - \\  
32 & 16:41:21 & +00:30:27 & 0.4533 & 0.654 &  97 & 1.87 (30) & V502\,Oph; WU \\ % ASAS2
33 & 18:41:39 & -00:44.7 & 0.2875 & 0.923 & 205 & 1.26 (30) & - \\  % ASAS2
34 & $^{*}$20:48:59 &  00:27.4 & 0.5134 & 0.664 & 493 & 2.90 (30) \\ % ASAS2
35 & 22:02:48 & -12:18.7 & 0.3067 & 0.873 & 173 & 1.02 (30) & - \\ % ASAS2
36 & 23:24:16 & -62:22:06 & 0.3577 &  0.71 & 136 & 9.68 (29) & BV\,1006 \\ 
\enddata
 \label{tab:1}
\end{deluxetable}

\begin{deluxetable}{cccccl}
\tablewidth{0pt}
\renewcommand{\arraystretch}{.6}
%\tabletypesize\footnotesize
\tablenum{1}
\pagestyle{empty}
\tablecaption{Comparison of Distance Determinations }
\tablehead{
 \colhead{RA} & \colhead{DEC} & \colhead{Dist.$_{\rm W UMa}$ }  &
 \colhead{$\pi$ ($\sigma_{\pi}$)} &\colhead{ Dist.$_{\it HIPPARCOS}$ } 
 & \colhead{ID }
          }
\startdata
02:38:33 & -14:17:54 & 107 &  3.82 (2.17) & 262 (167--606) & DY~Cet; HD\,16515 \\
04:12:09 & -10:28:12 &  67 & 17.96 (1.2)  & 56 (52--60)   & YY~Eri; HD\,26609 \\
04:33:25 & -23:56:18 & 228 &  2.21 (1.13) & 452 (299--926) & HN~Eri; HD\,29053\\
\enddata
 \label{tab:2}
\end{deluxetable}

%%%%%%%%%%%%%%%%%%%%%%%%%%%%%%%%%%%%%%
\section{X-ray Activity and Rotation}
%%%%%%%%%%%%%%%%%%%%%%%%%%%%%%%%%%%%%%

\begin{figure}
 \plotone{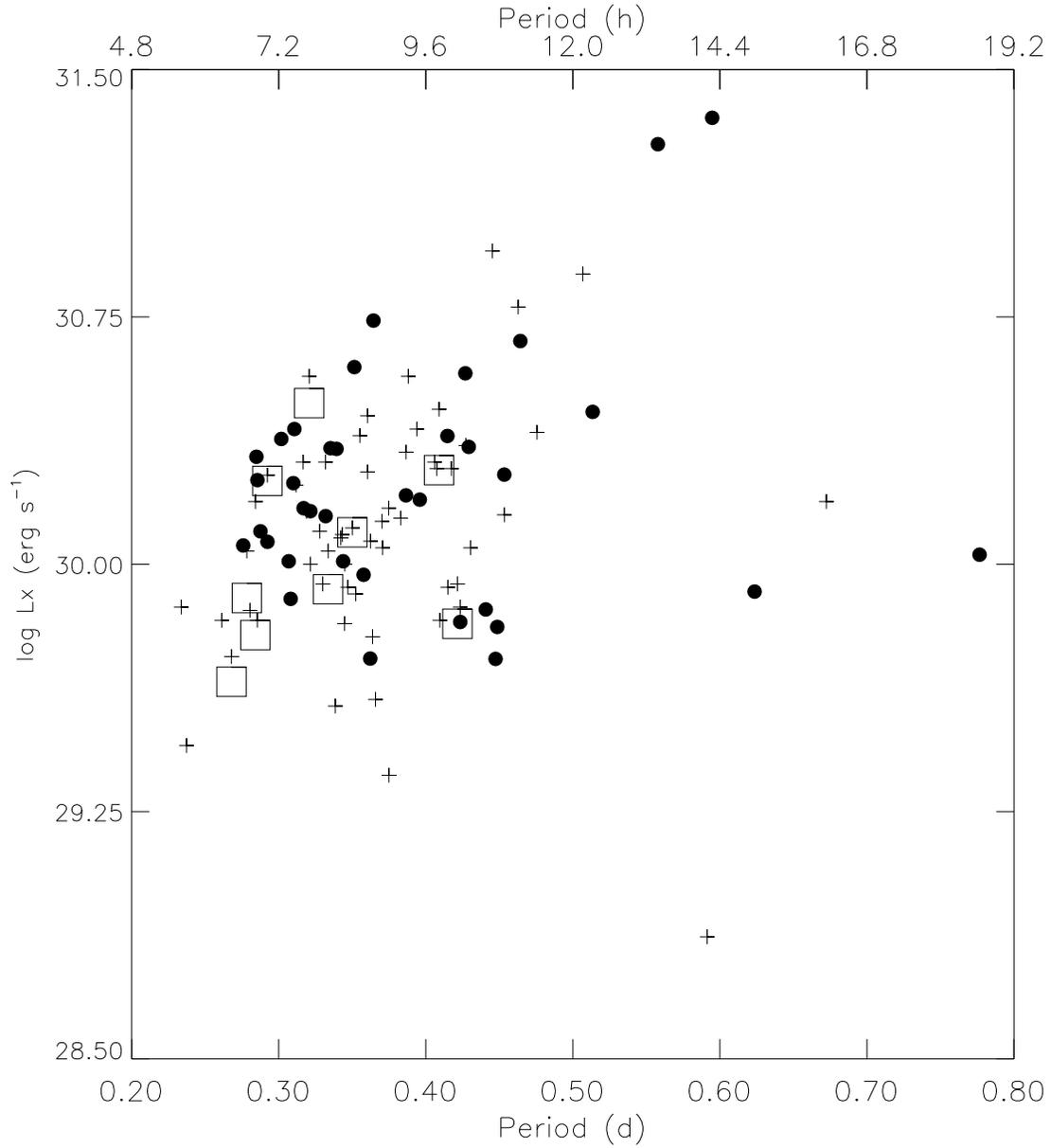}
  \caption{X-ray luminosity versus rotational period for W~UMa stars.  
          The solid circles are data from this work (Table~1).  Crosses  
          are data taken from \citet{ste01} and squares are data taken from 
         \citet{mcg96}. 
      }
  \label{fig:plx}
\end{figure}

The X-ray luminosity is known to increase with rotation in late-type field 
and cluster dwarfs \citep{piz03, pal81}, which is attributed to
their enhanced dynamo magnetic activity.  The trend holds until the rotation 
is faster than the period $P \leq 1$~d (Fig.~\ref{fig:plx}) at which saturation occurs.    

The W~UMa stars are tidally locked fast rotators, with the majority of 
periods shorter than 0.63~d.  Because of their nearly edge-on orbital 
orientation, W~UMa stars offer a good chance 
to investigate the relationship between rotation and magnetism, not only  
in a contact binary environment, but in a fast rotating system in general.    
The X-ray luminosities of the W~UMa stars in our sample (Table~\ref{tab:1}) range from
$ 4.37 \times 10^{29}$ to $2.26 \times 10^{31}$~erg~s$^{-1}$.  

It has been an unsettled issue whether the X-ray luminosity $L_{x}$ or 
its ratio to the total stellar luminosity $L_{\rm X} / L_{bol}$ should be
used as a measure of stellar X-ray activity.  The X-ray luminosity is directly 
proportional to the emitting volume, whatever the radiation mechanism, 
whereas the ratio compares the X-ray with photospheric emission.  
The situation is particularly unclear for W~UMa stars for which 
two stars share a single common envelope.  Likewise, different parameters have 
been used in the literature to quantify stellar rotation, e.g., 
the rotational period, equatorial speed ($ v \sin i$), or the Rossby number.   
Fig.~\ref{fig:plx} plots the $L_{\rm X}$ versus period for W~UMa stars in our
sample, together with those in \citet{ste01} and \citet{mcg96}.
One sees that the addition of our sample reinforces 
the notion that, for a rotational period shorter than $\sim 0.5$~d), 
the faster a W~UMa star rotates, the weaker 
its X-ray emission.  Such "supersaturation" has been 
suggested already in early study of W~UMa stars by {\it Einstein Observatory} 
\citep{cru84}, and lately discussed by \citet{ste01}. 
A similar anti-correlation between the X-ray luminosity 
and rotation might also exist for single stars with rotation faster 
than a saturation value ($\la 1$~d, \citet{piz03}; 
$v \sin i \ga 100$~km~s$^{-1}$, \citet{pro96}).  

\begin{figure}
 \plottwo{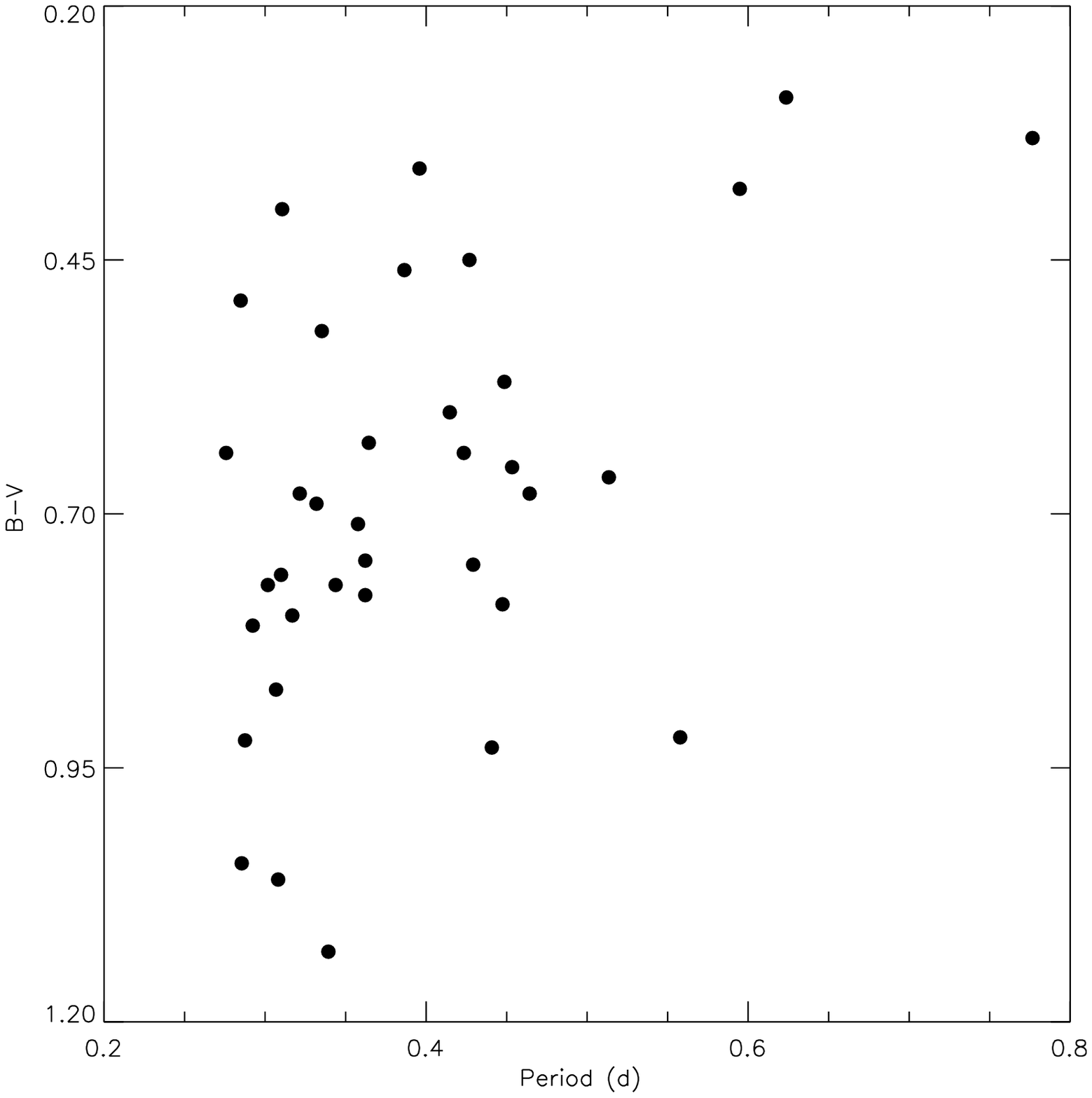}{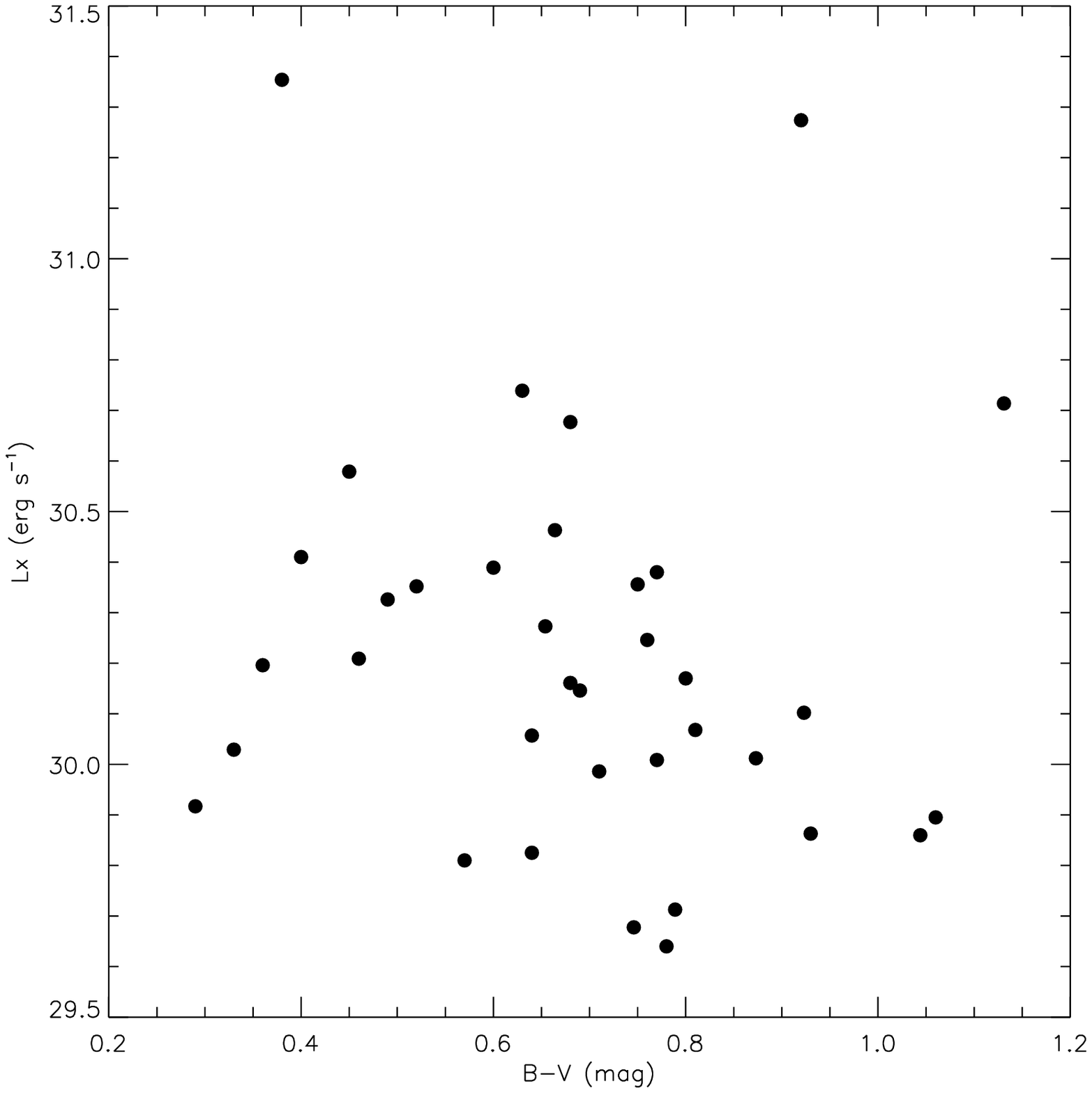}
  \caption{(Left) The observed (B-V) color versus period and (right) 
          the $L_{\rm X}$ versus $(B-V)$ for our sample stars. 
      }
  \label{fig:bv}
\end{figure}

Fig.~\ref{fig:bv} plots the the {\it observed} $(B-V)$ color 
versus the period of our sample stars.  In general the reddening 
$E(B-V)$ is small for a nearby ~W~UMa star \citep{ruc97b}, as is the case
for stars in our study.  Assuming the observed $(B-V)$ is close to 
the dereddened $(B-V)_{0}$, Fig.~\ref{fig:bv} resembles that shown   
by \citet{ruc97b}, so one cannot rule out the apparent (anti)correlation 
between X-ray luminosity and rotational period via the dependence 
of both quantities on the color.  The dependence of $L_{x}$ on 
$(B-V)$, however, is weak and scattered.   
 
In summary, we have identified 36 stars in the ASAS variable star 
database which are also X-ray sources.  Our sample substantially expands  
the list of known W~UMa stars as fast rotators which show supersaturation 
phenomenon, in the sense of decreasing X-ray luminosity as 
the rotation gets faster.  

\acknowledgments 
We thank Bahdan Paczinski for bringing up our attention to the issue of 
variable stars as X-ray sources.  
We also acknowledge Ron Taam for useful discussions.  This research is partially
supported by the grant NSC94-2112-M-008-017 of National Science Council.    

% The time scales of merging of the binary components in a W~UMa system 
% is short, some half billion years \citep{dry02}, compared to  

% Fig. 5 X-ray luminosity vs. rotation of field dwarfs (crosses) and 
% cluster stars (squares). Leftward arrows indicate field stars with periods 
% derived from {\it v sini} data, taken from \cite{piz03}.

%---------------------References-------------------------------

\acknowledgments

\end{document}